\documentclass[preprint, double,showpacs,preprintnumbers,amsmath,amssymb,
superscriptaddress,nofootinbib,longbibliography]{revtex4-1}
\usepackage{booktabs}
\usepackage{epsfig}
\usepackage{graphicx}
\usepackage{color}
\usepackage{hyperref}
\usepackage{amsmath}
\usepackage{tikz-feynman}
\usepackage{wrapfig}
\usepackage{cases}
\usepackage[utf8]{inputenc}

\usepackage{stackengine}
\stackMath
\usepackage{graphicx}

\newcommand{\pf}{\textrm{Pf}}

\AtBeginDocument{
\heavyrulewidth=.08em
\lightrulewidth=.05em
\cmidrulewidth=.03em
\belowrulesep=.65ex
\belowbottomsep=0pt
\aboverulesep=.4ex
\abovetopsep=0pt
\cmidrulesep=\doublerulesep
\cmidrulekern=.5em
\defaultaddspace=.5em
}

\begin{document}
\title{UV consistency conditions for CHY integrands}
\author{Laurentiu Rodina}
\affiliation{Institut de Physique Theorique, Universite Paris Saclay, CEA, CNRS, F-91191 Gif-sur-Yvette, France}
\date{\today}

\begin{abstract} 
We extend on-shell bootstrap methods from spacetime amplitudes to the worldsheet objects of the CHY formalism. We find that the integrands corresponding to tree-level non-linear sigma model, Yang-Mills and $(DF)^2$ theory are determined by demanding enhanced UV scaling under BCFW shifts. Back in spacetime, we also find that $(DF)^2$ theory is fixed by gauge invariance/UV scaling and simple locality assumptions.
\end{abstract}
\maketitle

\newpage
\section{Introduction}
The S-matrix bootstrap program aims to construct scattering amplitudes directly from on-shell physical principles, foregoing the complicated formalisms of off-shell Lagrangians or Feynman diagrams. In this approach, the BCFW recursion and generalized unitarity proved immensely powerful tools, enabling the calculation of otherwise intractable tree and loop level amplitudes \cite{BCFW,UnitarityMethod}, ultimately culminating in the purely geometric description of amplitudes in terms of geometries \cite{Arkani-Hamed:2013jha,Arkani-Hamed:2017mur}, while other advances uncovered a remarkable inter-connectivity between amplitudes: the double copy \cite{BCJ,BCJLoop,bcjReview,Carrasco:2019yyn}, transmutation operators \cite{CheungUnifyingRelations}, and most recently entanglement \cite{Cheung:2020uts}.

At the same time it has been discovered that amplitudes can be fully determined with even less information than previously thought: gauge invariance or the Adler zero \cite{AdlerZero, Cheung:2014dqa,Cheung:2016drk,Arkani-Hamed:2016rak,Rodina:2016jyz}, BCJ relations \cite{Carrasco:2019qwr}, IR behavior \cite{RodinaSoft} or UV scaling \cite{Rodina:2016mbk,Carrasco:2019qwr}, together with simple locality assumptions, are each sufficient to fully constrain a wide range of tree level amplitudes, including their unitarity (factorization). 

Of such constraints, UV scaling has surprisingly proven to be the most versatile and powerful: almost all theories have some hidden enhanced scaling under either single or two particle BCFW shifts, and for some of them this seems to be a fundamental and defining property:
\begin{center}
\begin{tabular}{ |c|c|c| } 
 \hline
  & adjacent shift& non-adjacent shift \\ \hline
 Yang-Mills & $z^{-1}$ & $z^{-2}$ \\ \hline
 Gravity &  & $z^{-2}$ \\ \hline
Born-Infeld &  & $z^0$\\ \hline
NLSM & $z$ & $z^0$ \\ \hline
special Galileon &  & $z^2$\\ \hline
\end{tabular}
\end{center}
\vspace{10pt}

The original motivation for studying this UV scaling was as a pre-condition for the BCFW recursion \cite{BCFW}. In this construction the momenta of some particles are deformed by a BCFW shift $p\rightarrow p+z q$, and if the amplitude has a scaling of  $\mathcal{O}(z^{-1})$ or better as $z \rightarrow \infty$, this implies the amplitude has only finite poles in $z$, and so may be rebuilt recursively from its residues, which are products of lower point amplitudes. Gravity exhibits and even better scaling of $\mathcal{O}(z^{-2})$  \cite{ArkaniHamed:2008yf,ArkaniHamed2008gz,Schuster:2008nh,McGady:2014lqa}, which leads to ``bonus relations" \cite{Spradlin:2008bu}. On the other hand it has recently become apparent that a recursion may be set-up even with poles at infinity, as long as the asymptotic behavior is sufficiently tame and can be probed by other properties  \cite{Kampf:2013vha,Kampf:2012fn,Cheung:2015cba,Luo:2015tat,Cachazo2016njl,Low:2019ynd,Elvang:2018dco}, while other UV and unitarity considerations may fix gravity loop integrands \cite{Edison:2019ovj}. However, the above table demonstrates that the UV scaling may be considered more fundamental than the recursion itself, as no other input beyond locality is needed. Unfortunately conditions like UV scaling, or even simple gauge invariance, are quite non-transparent as they require numerous cancellations between different Feynman diagrams, but if they are completely constraining this suggests that scattering amplitudes may have radically different descriptions, which make other properties manifest, at the expense of the traditional locality and/or unitarity.

A step in this direction is given by the CHY formalism \cite{Cachazo:2013gna,Cachazo:2013hca,Cachazo:2013iea,Cachazo2014nsa,Cachazo2014xea,He:2016iqi,Azevedo:2017lkz,Dolan:2013isa}, which can be viewed as a transform from spacetime to a worldsheet, where local singularities corresponding to propagators are replaced by worldsheet singularities $(\sigma_i-\sigma_j)^{-1}$, with the precise map encoded in the {\em scattering equations} (SE):
\begin{align}
A=\int d\sigma \delta(\textrm{SE}) I_L I_R
\end{align}
Manifest locality and unitarity inherent to Feynman diagrams are lost, but  gauge invariance and the double copy are more transparent in this formalism. This construction can be traced to an ambitwistor string theory \cite{Mason:2013sva,Casali2015vta}, but it is not clear what (if any) principles determine the worldsheet objects $I$ directly.

In this article we propose that simple scaling under BCFW shifts is sufficient to fully determine the integrands relevant for NLSM, YM, as well as $DF^2$ theory \cite{Johansson2017srf}. The only assumption needed is that these integrands are permutation invariant functions of dot products between momenta and polarization vectors (with no $(e\cdot  e)$ factors for $DF^2$ theory, a necessary extra assumption), with the only singularities coming from products of $\sigma_{ij}=\sigma_i-\sigma_j$, initially unrelated to the numerators. This result holds even without assuming the scattering equations, but we do require that shifting particles $i$ and $j$ also shifts the corresponding $\sigma$'s:
\begin{align}
\sigma_i&\rightarrow  \sigma_i+z \sigma_i\, ,\\
 \sigma_j&\rightarrow \sigma_j+z \sigma_j\, .
\end{align}
Under such shifts, we find that all three integrands scale as\footnote{We note that some extra care is required when discussing permutation invariance and UV scaling for NLSM and YM, which are given by reduced Pfaffians. As will be discussed later, for these integrands, two rows/columns are deleted - we associate these deletions with two particles which behave differently from the others, since we are not using the scattering equations: the integrands are not permutation invariant in these particles, and scalings involving them can be slightly worse.}:
\begin{align}
\mathcal{O}(z^{-2})
\end{align}

Next we can also consider what happens when we assume ``worldsheet locality": demanding that any dot product $x_i\!\cdot\!  y_j$ is rescaled only by the appropriate worldsheet factor $(\sigma_{ij})^{-1}$.  This provides an ansatz that is very close to manifesting the correct scaling: for vector theories, no term scales worse than $\mathcal{O}(z^{-1})$, so only a minor improvement is required to obtain the correct scaling of $\mathcal{O}(z^{-2})$. Moreover, the correct scaling in most (but not all) shifts holds term by term in the expansion of the Yang-Mills reduced Pfaffian, while the cycle expansion of the non-reduced Pfaffian manifests correct scaling term by term. This suggests that the worldsheet is in fact a more natural home for the BCFW shift, getting us closer to a formalism which trades locality and/or unitarity for enhanced UV behavior.

The paper is organized as follows. In Section~\ref{sec2} we first briefly review the CHY construction (a full description can be found in refs.~\cite{Cachazo:2013gna,Cachazo:2013hca,Cachazo:2013iea,Cachazo2014nsa,Cachazo2014xea}). In Sections~\ref{sec3}-\ref{sec5}  we then present evidence that the integrands relevant for NLSM, YM and $DF^2$ theory are fully determined by UV scaling and various assumptions. Motivated by this result, we also extend such observations to the usual spacetime  $DF^2$ amplitudes, which we find to be fixed by gauge invariance and similar UV conditions. We conclude with possible future directions in Section~\ref{secf}.

\section{CHY review}\label{sec2}
The CHY formula expresses various scattering amplitudes as:
\begin{align}
\label{chyint}
A_n=\int d\Omega \delta(\textrm{SE}) I_L I_R\, ,
\end{align}
where the scattering equations are given by:
\begin{align}
\textrm{SE}_i=\sum_{j\neq i}^n \frac{p_i\!\cdot\!  p_j}{\sigma_i-\sigma_j}\, ,\quad i=\overline{1,n}
\end{align}
and the half-integrands $I$ are functions of kinematics, polarization vectors (for vector theories), and worldsheet coordinates $\sigma$. Their specific expression dictates the particular theory to be obtained, in a form manifesting the BCJ double-copy. There are four ingredients relevant to our discussion, necessary to build NLSM, YM and $DF^2$ theory. For all three theories, one of the integrands is the Parke-Taylor factor which encodes the ordering of the resulting amplitudes:
\begin{align}
I_\textrm{PT}(1,2,\ldots,n)=\frac{1}{(\sigma_1-\sigma_2)(\sigma_2-\sigma_3)\ldots(\sigma_n-\sigma_1)}\, .
\end{align} 

The NLSM integrand is given by:
\begin{align}\label{nlsmpf}
I_\textrm{NLSM}=(\pf\, A^{ab})^2=\textrm{det}(A^{ab})\, ,
\end{align}
where the reduced matrix $A^{ab}$ is obtained by removing rows and columns $a$ and $b$ from the $n\times n$ matrix $A$:
\begin{align}
A_{ij}=\frac{p_i\!\cdot\!p_j}{\sigma_{ij}}\, .
\end{align}
In this paper we will focus on the simpler object $\sqrt{I_{\textrm{NLSM}}}=\pf\, A^{ab})$.

The YM integrand is given by:
\begin{align}
I_\textrm{YM}=\pf\, {\bf\Psi}^{ab}\, ,
\end{align}
where the reduced matrix ${\bf \Psi}^{ab}$ is obtained by removing rows and columns $a$ and $b$ from the $2n\times 2n$ matrix ${\bf \Psi}$:
\begin{align}
{\bf \Psi }=\begin{pmatrix}A &-C^{T} \\ C & B\end{pmatrix}\, ,
\end{align}
with
\begin{align}
A_{ij}=\frac{p_i\!\cdot\!p_j}{\sigma_{ij}},\qquad B_{ij}=\frac{e_i\!\cdot\!e_j}{\sigma_{ij}},\qquad C_{ij}=\frac{e_i\!\cdot\!p_j}{\sigma_{ij}}\, .
\end{align}
Despite this reduction, both integrands are permutation invariant on the support of the scattering equations. We will keep track of the labels $a$ and $b$, and associate them with two particles which we single out as having a distinct scaling behavior under shifts. To avoid clutter, we will sometimes drop the extra label $ab$ from objects under consideration.

The $DF^2$ integrand is given by:
\begin{align}
I_{DF^2}=\prod_{i=1}^n \sum_{j\neq i}^n \frac{e_i\!\cdot\!  p_j}{\sigma_{ij}}\, ,
\end{align}
and is directly permutation invariant. Note that the only dot products appearing are of the form $(e \!\cdot\!  p)$, a fact we use as an assumption.

Other theories such as gravity, Born-Infeld or the special Galileon may be obtained by mixing these ingredients:
\begin{align}
\textrm{GR} &\sim I_{\textrm{YM}} I_{\textrm{YM}}\, ,\\
\textrm{BI}& \sim I_{\textrm{YM}} I_{\textrm{NLSM}}\, ,\\
\textrm{sGal}& \sim I_{\textrm{NLSM}} I_{\textrm{NLSM}}\, .
\end{align}
It is important to note that the integral (\ref{chyint}) fully localizes on the delta functions, so in fact the amplitude is simply given by a sum over solutions to the scattering equations. 
\begin{align}
A_n=\sum_{\sigma^*}J(\sigma^*) I_L(\sigma^*) I_R(\sigma^*)\, ,
\end{align}
where $J$ is some Jacobian factor resulting from solving the delta functions. In practice the scattering equations are non-trivial to solve (see \cite{Baadsgaard:2015voa,Bjerrum-Bohr:2016juj,Zhou:2017mfj,Lam:2018tgm} for developments), but they do not enter into our discussion.

\subsection{BCFW scaling}
We will use the following BCFW two particle shift \cite{Rodina:2016mbk,Carrasco:2019qwr}:
\begin{align}
e_i&\rightarrow \hat{e}_i\, ,\\
e_j&\rightarrow \hat{e}_j+z p_i\frac{\hat{e}_i\!\cdot\!  e_j}{p_i\!\cdot\!  p_j}\, ,\\
p_i&\rightarrow p_i-z \hat{e}_i\, ,\\
p_j&\rightarrow p_j+z \hat{e}_i\, ,
\end{align}
where
\begin{align} \hat{e}_i=e_i-p_i\frac{e_i\!\cdot\!  p_j}{p_i\!\cdot\!  p_j}\, .
\end{align} 
We also rescale in the worldsheet coordinates:
\begin{align}
\sigma_i&\rightarrow  (1+z)\sigma_i\, ,\\
\sigma_j&\rightarrow  (1+z)\sigma_j\, .
\end{align}
Under this shift  we find the following scalings:\\
 for NLSM, $\pf\, A^{ab}$:
 \begin{itemize}
 \item $[i,j\rangle \sim \mathcal{O}(z^{-1})$
 \item $[i,a\rangle,[i,b\rangle,[a,b\rangle\sim \mathcal{O}(z^0)$
 \end{itemize}
 for YM, $\pf\, {\bf\Psi}^{ab}$:
  \begin{itemize}
 \item $[i,j\rangle,[i,a\rangle,[i,b\rangle,[a,i\rangle,[b,i\rangle \sim \mathcal{O}(z^{-2})$
 \item $[a,b\rangle,[b,a\rangle \sim \mathcal{O}(z^{-1})$
 \end{itemize}
 and for $DF^2$:
  \begin{itemize}
 \item $[i,j\rangle \sim \mathcal{O}(z^{-2})$
 \end{itemize}

The $\sigma$ shifts can be motivated by requiring the scattering equations corresponding to particles $i$ and $j$ to have an improved scaling under the combined shift \cite{Dolan:2013isa}:
\begin{align}
\textrm{SE}_i&=\sum_k \frac{p_i\!\cdot\!  p_k}{\sigma_{ik}} \rightarrow \mathcal{O}(z^{-1})\, , \\
\textrm{SE}_j&=\sum_k \frac{p_j\!\cdot\!  p_k}{\sigma_{jk}} \rightarrow \mathcal{O}(z^{-1})\, .
\end{align}
Note that the other SE do not have this behavior. For instance, at 4 points under a $[1,2\rangle$ shift, the following scattering equation becomes:
\begin{align}
\textrm{SE}_3=\sum_k \frac{p_3\!\cdot\!  p_k}{\sigma_{3k}} \longrightarrow z^0\left(\frac{e_1\!\cdot\!  p_4}{\sigma_1}-\frac{e_1\!\cdot\!  p_4}{\sigma_2}+\frac{p_1\!\cdot\!  p_2}{\sigma_3-\sigma_4}\right)+\mathcal{O}(z^{-1})\, .
\end{align}
While not mandatory, solving these shifted equations up to order $\mathcal{O}(z^{-2})$ further improves the scaling of the YM integrand to $\mathcal{O}(z^{-4})$.

It is interesting to note that the Laplace expansion of the reduced Pfaffian,
\begin{align}
\pf\, M=\sum (-1)^{i+j...} m_{ij}\pf\, M^{ij}\, ,
\end{align}
manifests the right BCFW scaling term by term in all shifts except those involving the ``reduced" particles. Similarly, the cycle expansion of the full Pfaffian (which sums to zero)  \cite{He:2016iqi,Lam:2016tlk}, manifests both correct scaling and permutation invariance term by term:
\begin{align}
\pf\, {\bf\Psi}_n=\sum (-1)^{n-m} P_{i_1 i_2\ldots i_m}\, .
\end{align}
These special building blocks are given by:
\begin{align}
\label{block}
P_{i_1i_2\ldots i_r}=\sum_{|I_1|=i_i, |I_2|=i_2,\ldots|I_r|=i_r} \Psi_{I_1}\Psi_{I_2}\ldots\Psi_{I_r}\, ,
\end{align}
where the $\Psi$'s are gauge invariant, either via linearized field strengths or via the scattering equations:
\begin{align}
\Psi_{(a)}&=-\sum _{j\neq a} \frac{e_a\!\cdot\!  p_j}{\sigma_{aj}}\\
\Psi_{(a_1a_2\ldots a_3)}&=\frac{\frac{1}{2}\textrm{tr}(f_{a_1}f_{a_2}\ldots f_{a_i})}{\sigma_{a_1a_2}\sigma_{a_2 a_3}\ldots \sigma_{a_i a_1}},\qquad \textrm{with } f_a^{\mu\nu}=p_a^\mu e_a^\nu- e_a^\mu p_a^\nu
\end{align}
Note that in this notation the integrand for $DF^2$ is precisely $P_{1111\ldots 1}$.
\subsection{Worldsheet locality}
One property common to all integrands is what we will call worldsheet locality: the correspondence between worldsheet coordinates and spacetime dot products:
\begin{align}
\frac{x_i\!\cdot\!  y_i}{\sigma_i-\sigma_j}\, .
\end{align}
With integrands given as a sum over products of such factors. This has an immediate effect on UV scaling: no factor may scale worse than $\mathcal{O}(z^0)$, while for vector theories, because of multilinearity in polarization vectors, no term can scale worse than $\mathcal{O}(z^{-1})$.

Quite surprisingly, we will see that even this property follows directly from demanding improved scaling, as our initial ansatze allow ``non-local" factors:
\begin{align}
\frac{x_i\!\cdot\!  y_i}{\sigma_k-\sigma_l}\, ,
\end{align}
where $k$ and $l$ need not be related to $i$ and $j$.
\section{Non-linear sigma model}\label{sec3}
As mentioned before, we are looking at essentially $\sqrt{I_\textrm{NLSM}}$. Therefore our ansatz is defined by:
\begin{itemize}
\item numerators are  $(n/2-1)$ dot products of $p_i\!\cdot\!  p_j$
\item denominators are $(n/2-1)$ factors of $\sigma_{ij}$.
\end{itemize}
For instance, the four point ansatz is given by:
\begin{align}
B_4=\sum_{i,j,k,l} a_{ijkl} \frac{p_i\!\cdot\!  p_j}{\sigma_{kl}}\,.
\end{align}
Then after fixing some $a$ and $b$, we impose the following scalings:
\begin{align}
[i,j\rangle & \sim \mathcal{O}(z^{-1})\, ,\\
[i,a\rangle ,[i,b\rangle, [a,b\rangle &\sim \mathcal{O}(z^{0})\,,
\end{align}
for all $i,j\neq a,b$. In the four point case, choosing $a=3$ and $b=4$ only the $[1,2\rangle$ shift must scale as $\mathcal{O}(z^{-1})$, the others shifts must scale as $\mathcal{O}(z^0)$. Imposing this behavior we find a unique solution:
\begin{align}
\pf\, A^{34}_4=\frac{p_1\!\cdot\!  p_2}{\sigma_{12}}\,.
\end{align}
We can also perform the check at 6 points:
\begin{align}
B_{6}=\sum \frac{(p\!\cdot\!  p)^{2}}{(\sigma_{ij})^{2}}\, ,
\end{align}
this ansatz already has around 5000 terms, but imposing the correct scaling, with $a=5$ and $b=6$, we find a unique solution:
\begin{align}
\pf\, A^{56}_6=\frac{p_1\!\cdot\! p_4\, p_2\!\cdot\! p_3}{\sigma _{14} \sigma _{23}}-\frac{p_1\!\cdot\! p_3\, p_2\!\cdot\! p_4}{\sigma _{13} \sigma _{24}}+\frac{p_1\!\cdot\! p_2\, p_3\!\cdot\! p_4}{\sigma _{12} \sigma _{34}}\, .
\end{align}

\section{Yang-Mills}\label{sec4}
We are using the following assumptions:
\begin{itemize}
\item numerators are multi-linear in all polarization vectors, with mass dimension $[n-2]$
\item denominators are products of $(n-1)$ $\sigma_{ij}$ factors
\item permutation invariance in $(n-2)$ particles
\end{itemize}
At four points such an ansatz looks like:
\begin{align}
B_4=a_1 \frac{e_1\!\cdot\!   e_2\, e_3\!\cdot\!  p_2\, e_4\!\cdot\!  p_2}{\sigma_{12}\sigma_{13}\sigma_{14}}+\ldots\, ,
\end{align}
where again we note that the denominators are unrelated to the numerators. We then claim that the YM reduced Pfaffian of ${\bf \Psi}^{ab}$ is uniquely fixed by the following scalings under BCFW shifts:
\begin{align}
[i,j\rangle,[i,a\rangle ,[i,b\rangle \sim \mathcal{O}(z^{-2})\, ,\\
[a,b\rangle, [b,a\rangle \sim \mathcal{O}(z^{-1})\,.
\end{align}
which is easily checked at four points.

At five points the ansatz is already too large to verify the claim, but there is a simple assumption which reduces the complexity: that each $\sigma_i$ should appear at least once per term. Then we are able to find a unique solution: the reduced five point Yang-Mills Pfaffian.

\section{$DF^2$ theory} \label{sec5}
We are using the following assumptions:
\begin{itemize}
\item numerators are multi-linear in all polarization vectors, with mass dimension $[n]$, no $(e\!\cdot\!  e)$ dot products allowed
\item denominators are products of $n$ $\sigma_{ij}$ factors
\item permutation invariance in $n$ particles
\end{itemize}
So an ansatz looks like:
\begin{align}
B_4=a_1 \frac{e_1\!\cdot\!   p_2\, e_2\!\cdot\!  p_1\, e_3\!\cdot\!  p_2 \, e_4\!\cdot\!  p_2}{\sigma_{12}\sigma_{13}\sigma_{14}\sigma_{34}}+\ldots\, .
\end{align}
Then we claim that the $DF^2$ integrand is uniquely fixed by a BCFW scaling of $\mathcal{O}(z^{-2})$ under all two particle shifts. This is easily verified at four points, but at five points we again need to impose the presence of every $\sigma_i$ in each term, which allows us to check a somewhat weaker claim.

It is interesting to note that if we also allow $(e\!\cdot\!  e)$ factors in the ansatz, we obtain new solutions which include all the $P_{i_1 i_2\ldots i_r}$  building blocks of the cycle expansion of the Pfaffian in eq. \ref{block}. At four points we only obtain one extra solution, which can be eliminated either by imposing gauge invariance (on the support of the scattering equations), or by demanding $\mathcal{O}(z^1)$ scaling under the transformation:
\begin{align}
p_i&\rightarrow z p_i\, ,\\
\sigma_i&\rightarrow z \sigma_i\, .
\end{align} 
It remains an open question whether this is still true at higher points, and whether there exist any constraints which select the linear combinations of the $P_{i_1 i_2\ldots i_r}$ that correspond to $F^3$-type interactions \cite{He:2016iqi}.

\subsection{Spacetime $DF^2$}
In spacetime we are back to usual propagators, which in the case of $DF^2$ \cite{Johansson2017srf}, a non-unitary theory, come in two types: gluon propagators, which are sum of consecutive momenta  $(\sum p_i)^4$, and scalar propagators, which can be sums of non-consecutive momenta $(\sum p_i)^2$, since the scalar color structure allows non-planar interactions. Like the CHY integrand, we do not allow dot products $(e \!\cdot\!  e)$, but only $(e\!\cdot\!  p )$ and $(p \!\cdot\!  p)$. A four point ansatz is given by:
\begin{align}
A_4=\frac{n_s(p^6)}{s^2}+\frac{n_t(p^6)}{t^2}+\frac{n_u(p^4)}{u}\,.
\end{align}
Imposing gauge invariance in all four particles:
\begin{align}
e_i\rightarrow p_i \Rightarrow A_4=0\,,
\end{align}
we find a unique result, the $DF^2$ amplitude. This is similar to the YM uniqueness from gauge invariance \cite{Arkani-Hamed:2016rak,Rodina:2016jyz}. Note that the numerators cannot be independently gauge invariant in all particles because even if they have a sufficiently high mass dimension, they are only functions of $e\!\cdot\!  p$ and $p\!\cdot\!  p$. A polynomial of the form $f_n((e \!\cdot\!  p)^n,(p\!\cdot\!  p)^k )$ may only be gauge invariant in $k$ particles, while $DF^2$ numerators are of the form $f_n((e\!\cdot\!  p)^n, (p\!\cdot\!  p)^{n-3})$. Therefore only a sum of diagrams can be fully gauge invariant, again similar to Yang-Mills.

Finally, we also find that the BCFW scalings:
\begin{align}
\mathcal{O}(z^1)& \textrm{  for adjacent shifts}\,,\\
\mathcal{O}(z^0)& \textrm{  for non-adjacent shifts}\,,
\end{align}
uniquely fix the $DF^2$ amplitudes, at least up to $n=5$.

\section{Conclusion}\label{secf}
We have shown that the constraining power of BCFW shifts is sufficient to fully determine the main ingredients of the CHY formalism, further demonstrating the surprising universality of enhanced UV scaling. While the claim for YM and $DF^2$ theory only holds if also assuming permutation invariance, it appears that this assumption can be replaced with another type of UV scaling: single particle shifts $[i\rangle$, introduced in \cite{Carrasco:2019qwr}. These shifts are simply given by:
\begin{align}
e_i&\rightarrow \hat{e}_i\, ,\\
p_i&\rightarrow p_1-z \hat{e}_i\, ,\\
\sigma_i&\rightarrow  z\sigma_i\, .
\end{align}
At four points, we find that the usual two particle shifts, together with the single particle scalings:
\begin{align}
\textrm{YM}&: [i\rangle \sim \mathcal{O}(z^{n-5}) \textrm{ and } [a\rangle,[b\rangle \sim \mathcal{O}(z^{n-4})\,,\\
DF^2&: [i\rangle \sim\mathcal{O}(z^{n-3})\,
\end{align}
are enough to fully constrain the respective integrands, pointing to a description purely in terms of UV scalings.

The $\mathcal{O}(z^{-2})$ scaling also guarantees the lack of a pole at infinity, and so the integrands may be rebuilt from finite residues \footnote{The BCFW reconstruction of CHY was investigated in the original proof of this formalism in ref.~\cite{Dolan:2013isa}, but factorization based on the existence of scattering equations was crucial in that approach.}. It is worth mentioning that the $\mathcal{O}(z^{-2})$ of the integrands also implies the existence of so-called ``bonus relations" \cite{Spradlin:2008bu} between residues.

 A related theory is the so-called ``$DF^2$+YM", relevant to the bosonic string, and which also has a CHY construction \cite{He:2018pol,Azevedo2018dgo,He:2019drm}. It would be interesting to investigate whether there exist (purely field theoretic) consistency conditions which determine this object as well.

\section*{Acknowledgements} The author would like to thank John Joseph Carrasco, Song He and Yong Zhang for discussions, and is grateful to the Institute of Theoretical Physics, Chinese Academy of Sciences, Beijing, where this work was initiated. The author is supported by the European Research Council under ERC-STG-639729, {\it Strategic Predictions for Quantum Field Theories\/}. 

\bibliography{UVCHY}

\end{document}